\documentstyle[epsf,psfig,referee]{mn}

\begin{document}

\title[Anglo-Australian Planet Search]
{A probable planetary companion to HD~39091 from the Anglo-Australian Planet Search}

\author[H. R. A. Jones et al.]{
\parbox[t]{\textwidth}{Hugh R. A. Jones$^1$, R. Paul Butler$^2$, 
C.G. Tinney$^3$, Geoffrey W. Marcy$^4$,  
Alan J. Penny$^5$, Chris McCarthy$^2$,
Brad D. Carter$^6$, Dimitri Pourbaix$^7$} \\
\vspace*{6pt} \\
$^1$Astrophysics Research Institute, Liverpool John Moores University,
Egerton Wharf, Birkenhead CH41 1LD, UK\\
$^2$Carnegie Institution of Washington,Department of Terrestrial Magnetism,
5241 Broad Branch Rd NW, Washington, DC 20015-1305, USA \\
$^3$Anglo-Australian Observatory, PO Box 296, Epping. 1710, Australia\\  
$^4$Department of Astronomy, University of California, Berkeley, CA, 94720, USA\\
$^5$Rutherford Appleton Laboratory, Chilton, Didcot, Oxon OX11 0QX, UK\\
$^6$Faculty of Sciences,  University of Southern Queensland, Toowoomba, 
QLD 4350, Australia\\
$^7$Institut d'Astronomie et d'Astrophysique, Universite Libre de 
Bruxelles CP 226, Boulevard du Triomphe, B-1050 Bruxelles, Belgium\\
\\
}

\date{submitted to MNRAS}

\maketitle

\label{firstpage}

\begin{abstract}

We report the detection of a extrasolar planet
candidate orbiting the G1~V star HD 39091.  The
orbital period is 2083 d and the eccentricity  
is 0.62.  With a minimum (M~sin~$i$) mass of 10.3 M$_{\rm JUP}$
this object falls near the high mass end of the observed
planet mass function, and may plausibly be a brown   
dwarf.  Other characteristics of this system, including 
orbital eccentricity and metallicity, are typical of    
the well populated class of radial velocity planets     
in eccentric orbits around metal-rich stars.

\end{abstract}


\section{Introduction}
Since the discovery of the first extra-solar planet (Mayor \&
Queloz 1995), planetary detections have been dominated
by northern hemisphere search programmes -- most prolifically by the 
high precision velocity
programmes at Lick (e.g., Butler et al. 1996) and Keck (e.g. Vogt et al. 
2000) and lower precision
programmes at OHP (Baranne et al. 1996), McDonald Observatory (Cochran
et al. 1997), AFOE (Noyes et al. 1997), and programmes at La Silla 
(Kurster et al. 2000; Queloz et al. 2000).
Of these programmes, which achieve precisions of $\sim$ 10~m~s$^{-1}$,
only the latter has access to the sky south of $\sim -20\deg$, and
these achieve precisions of $\sim$ 10~m~s$^{-1}$. In 1998, the 
Anglo-Australian Planet Search (AAPS) began in the southern
hemisphere enabling all-sky coverage 
of the brightest stars at precisions reaching 3~m~s$^{-1}$.
In this paper we present the second set of results from this programme. 
Two companion papers, Tinney et al. (2002a, 2002b), present results for a 
further two planets and an initial investigation of the Calcium H and 
K activity among some of the target stars of the AAPS.  

Precision Doppler surveys have found all of the
known extrasolar planets around solar-type stars. 
Discoveries have included: the first system of
multiple planets orbiting a Sun-like star (Butler et al. 1999);
the first planet seen in transit (Henry et al. 2000,
Charbonneau et al. 2000); the
first two sub-Saturn-mass planets (Marcy, Butler \& Vogt 2000); and the AAPS'
discovery of the first planet in a circular orbit outside
the 0.1~au tidal-circularisation radius (Butler et al. 2001).
A number of major surprises have emerged from the
sample of extrasolar planets.  

\begin{itemize}

\item The sub-stellar companion mass
function for F, G, and K dwarfs rises strongly below
10 M$_{\rm JUP}$,  and shows no signs of flattening toward the detection limit
near 1 M$_{\rm JUP}$.  And surprisingly, brown dwarf companions to solar type
stars are rare (Butler et al. 2000). 
This is in spite of a strong selection bias in the
observations that makes brown dwarfs easier to detect than
planets. It should be noted that the mass is only known
within the uncertainty of the projection factor represented by
the sin~$i$ of the orbit. However, the sin~$i$ statistic is between
1 and 0.5 for 87 per cent of randomly inclined orbits, so 87 per cent
of reported $M$ sin $i$ values will be within a factor of two of the 
true masses.

\item About $\sim$0.75\% of nearby solar-type stars stars
have been found to have planets orbiting
in circularised orbits
inward of 0.1~au (the 51\,Peg-like ``hot
Jupiters''). A smaller fraction of stars is now 
being found to have ``eps Ret -like'' planets orbiting in circularised 
orbits at
1 au or so (Tinney et al. 2002). 
But the dominant class of extra solar planets, (found
around some 7 per cent of target stars) 
show highly eccentric orbits within 3.5 au.
None of these classes were predicted {\em a priori}
by planetary formation theories. Such theories have received enormous
impetus from these observations, leading to models for planet formation
and evolution which now include the effects of 
dynamical friction, disk-planet and planet-planet interactions
(e.g., Rasio \& Ford 1996; Weidenschilling \& Marzari 1996;
Artymowicz 2000; Boss 2000).

\item The majority of extrasolar planets that have been found so far
occur around stars which are mildy metal-rich ($\sim$+0.2 dex with
considerable scatter) compared to the Sun (Laughlin 2000; Santos, Israelian
\& Mayor 2002).

\end{itemize}

This paper reports the discovery of a new planet candidate from the 
AAPS. Section 2 describes this precision programme. The stellar properties
and Keplerian orbital fits for the new planet candidate are presented
in Section 3. Section 4 provides a discussion of the new object.

\section{The Anglo-Australian planet search}

The Anglo-Australian planet search (AAPS) is carried out on the 3.92m
Anglo-Australian Telescope using the University College London Echelle
Spectrograph (UCLES), operated in its 31 lines/mm mode 
together with an I$_{2}$ absorption cell before the slit. 
UCLES used to be operated with an MIT/LL 2048$\times$4096
15$\mu$m pixel CCD. Since August 2001 the AAPS has changed to 
the AAO's EEV 2048$\times$4096 13.5$\mu$m pixel CCD. This CCD provides
30\% better quantum efficiency across the 5000--6200~\AA~
region containing I$_{2}$ 
absorption lines. Furthermore, the EEVs smaller pixels
result in improved resolution and reduced charge diffusion.
This CCD change has not required any significant changes of observational
or reduction procedure and has resulted in significantly
increased signal-to-noise. In the future we envisage further improvements
in observing efficiency and S/N through the use of a (1) an exposure meter,
(2) a motorised system for moving the I$_{2}$ cell in and out of the beam
and (3) a prime focus fibre feed for UCLES. 

Doppler shifts are measured by observing through an I$_2$
cell mounted behind the UCLES slit. The resulting superimposed iodine
lines provide a fiducial wavelength scale against which to measure
radial velocity shifts. The shapes of the iodine lines convey the 
PSF of the spectrometer for changes in optics and illumination on 
all time scales. 
We synthesize the echelle spectrum of each observation on a sub-pixel grid
using a high resolution reference template, and fit for
spectrograph characteristics (the wavelength scale, scattered light
and the spectrograph PSF) and Doppler shift.
This analysis obtains velocities from multiple epoch observations, which
are measured against this reference template.
This reference template is an observation at the highest
available resolution (using a small 0.5 arcsec slit) and high
S$/$N = 200 to 300, without the I$_2$ cell present. Such measurements can only 
be efficently obtained
in good seeing and take about 4 times as long to acquire as a standard
epoch (I$_2$ and a 1 arcsec slit) observation.
We show in Fig. 1 the residuals about zero
velocity for a sample of apparently stable stars 
reported by Butler et al. (2001),
as well as the residuals about a Keplerian fit for several detected planets.
We achieve 3~m~s$^{-1}$ precisions
down to the V~=~7.5 magnitude limit of the survey.
The fundamental limit
to the precision that can be achieved for our sample is set by a combination
of S$/$N (which is dependent on good seeing and weather conditions),
and the intrinsic velocity stability of our target stars often
called ``jitter'', induced
by activity and/or star spots (Saar et al. 1998; Saar \& Fischer 2000),
rather than our observing technique.
There is currently no way to tell whether
a residual scatter of larger than 3~m~s$^{-1}$ is
due to a small-amplitude planet (either short- or long-period --
the detection of the latter is one of our primary goals, as
these are Jupiter-like signals), or jitter induced by
star spots and/or activity.
Only observations over a long enough period to allow the
search for long-term periodicites can reveal the presence of
such relatively small-amplitude long-period signals such as Jupiter.
It is therefore vital
to monitor all our targets
for the lifetime of the survey not just those that appear to
be good planet candidates.\\

Our target sample includes 178 FGK
stars with declinations below $\sim -20\deg$ and V$<$7.5, and a further 23 
metal-rich stars with V$<$11.5. Where age/activity information is
available from log $R_{\rm HK}$ indices (Henry et al. 1996; Tinney et al. 2002a)
we require target stars to have log $R_{\rm HK}$ $<$ --4.5 corresponding to
ages greater than 3 Gyr. Stars with known stellar companions within
2 arcsec are removed from the observing list, as it is operationally difficult
to get an uncontaminated spectrum of a star with a nearby companion.
Spectroscopic binaries discovered during the programme have also been 
removed and will be the subject of a separate paper
(Blundell et al. 2002, in preparation).
Otherwise there is no bias against observing multiple stars. The programme
is not expected to have any bias against brown dwarf companions. 
The observing and data processing procedures
follows that described in Butler et al. (1996,2001).
Our first observing run was in January 1998, and the last run for which
observations are reported here was in October 2001.
 
\section{Stellar Characteristics and Orbital Solution for HD~39091}

A total of 26 observations of HD~39091 (GJ 9189, HIP 26394, HR 2022) 
have been made by
HIPPARCOS yielding a distance of 18.2 pc and a V magnitude of 5.65 (ESA 1997).
The resulting absolute magnitude is $M_{\rm V}$ = 4.32.
The star is photometrically stable within the HIPPARCOS measurement error,
with a photometric scatter of 0.006 magnitudes. The Bright Star 
Catalogue (Hoffleit 1982)
assigns a spectral type of G1V, in reasonable agreement with the
HIPPARCOS spectral type of G3IV.  The star is chromospherically 
inactive with log $R_{\rm HK}$ = --4.97 (Henry et al. 1996).
Based on the spectroscopic analysis of Santos, Israelian \& Mayor (2002),
[Fe/H] = 0.09 and Mass = 1.1$\pm$0.02 M$_\odot$.
Decin et al. (2000) find no evidence for a 60 $\mu$m excess in 
deep ISO data for HD~39091.
It is interesting to note that Queloz et al. (2000) report 
HD~39091 as a non-variable star, however, the Queloz et al. 
survey would not yet have been sensitive to the long period
of HD~39091.

The 26 Doppler velocity measurements of HD~39091, obtained
between November 1998 and October 2001, are listed in Table 1 and  
shown graphically in Fig. 2. The third column labelled error
is the velocity uncertainty produced by our least-squares
fitting. This fit simultaneously determines the Doppler shift
and the spectrograph point-spread function for each observation
made through the iodine cell, given an iodine absorption spectrum
and an iodine-free template of the object (Butler et al. 1996).
This uncertainty includes the effects of photon-counting uncertainties,
residual errors in the spectrograph PSF model, and variation in
the underlying spectrum between the template and iodine epochs. All
velocities are measured relative to the zero-point defined by
the template observation. Only observations where the uncertainty
is less than twice the median uncertainty are listed.
The best-fit Keplerian curve
yields an orbital period of 2083 d, a velocity amplitude 
of 196 m~s$^{-1}$, and
an eccentricity of 0.62. The minimum (M~sin~$i$) mass of the
planet is 10.3 M$_{\rm JUP}$, and the semi-major axis is
3.34~au. The RMS to the Keplerian fit is 6.6 m~s$^{-1}$, 
yielding a reduced chi-squared of 1.1. The properties of 
the extra-solar planet in orbit around HD~39091 are summarised
in Table 2.
Whilst we do not yet have a full orbit for HD 39091, we
do clearly have a 2nd inflection, which means that
the orbit is starting to become clear. However,  
the parameters of the orbit are relatively
poorly constrained and any follow-up observations
should take this into account. We are announcing
this object at this somewhat early stage in order
that Adaptive Optics and astrometric
follow-up on this relative long-period
object can be started as soon as possible.

We investigated the HIPPARCOS dataset for HD~39091 
using the set of tests devised by Pourbaix \& Arenou (2001). 
These tests combine radial velocity measurements 
with those from HIPPARCOS to derive upper mass limits
for companions.  HD~39091 fails all the Pourbaix \& Arenou tests 
apart from one.  
The Thiele-Innes orbit does improve the fit. However,
the derived elements are not significantly different from zero 
and thus no constraint can be placed on the orbit.  The
reason for this is the long period.  It exceeds the mission duration
by a factor two so the size of the orbit is not constrained at all.
Neither does HIPPARCOS give a lower limit for the inclination.  
Such a limit is based on the amplitude (the astrometric semi-major 
axis of the orbit) corresponding to the spectroscopic period.  
Once again, the method can not be applied to HD~39091 since 
the period exceeds the mission duration.
Hence, unfortunately, HIPPARCOS cannot provide any additional 
information.

\section{Discussion \label{discussion}}

It is interesting to note the mass of
10.3 M$_{\rm JUP}$ places HD~39091's companion  
rather close to the canonical deuterium fusion 
limit of around 13 M$_{\rm JUP}$. Most definitions
of a planet posit that they should not be undergoing
any nuclear fusion. 
Unfortunately distinguishing the presence of 
deuterium and thus the lack of fusion is still beyond 
our current observational capabilities. Once 
we have (1) the capability to make high resolution and high
signal-to-noise infrared spectroscopic measurements of 
such faint close-in companions and (2) sufficiently high quality molecular
line lists of HDO and H$_2$O, HD~39091's mass and long orbit 
make it a prime target. 

The AAT, Keck and Lick precision velocity
surveys are currently surveying a total of 1200 stars. 
All three of these programmes use the Iodine
cell technique, and all three have demonstrated long-term
precision of 3m~s$^{-1}$ (Marcy \& Butler 1998; Vogt et al.
2000; Butler et al. 2001). 
The ellipticities for the substellar 
candidates from these surveys are shown in Fig. 3.  The extra-solar 
planets show a wide range of eccentricity 
similar to the eccentricity distribution of binary systems
with stellar companions (e.g., Duquennoy \& Mayor 1991). 
This is
not understood in terms of a global planetary formation
model that also allows for the almost circular orbits
in our Solar System. Given that we have only just 
started to explore the potential parameter space for
extra-solar planets, history should remind us not
to envisage the Solar System as a special case.
It is important to note that our 
studies have not yet run for long enough
to be sensitive to a true Solar System analogue.

The star HD~39091 is 
established as metal-rich star relative to the Sun
which is itself metal-rich compared to the average
of stars in the solar neighbourhood. The companion to HD~39091
reported here thus further confirms
the statistical findings of Santos
et al. (2002) that stars with planets are metal rich
compared to stars without planet detections. 
There are two general classes of explanation for this:
intrinsic metallicity bias and accretion of metal-rich
material. 
After a variety of different 
spectral analyses and claims a straightforward explanation 
remains: the higher metallicity of planet-harbouring stars 
arises because high metallicity environments 
have a higher probability of planet formation.
 
\section{Conclusions}

We present data showing evidence for an 
eccentric-orbit extra-solar
planet around the
metal-rich star HD~39091. The detection
of this long period object gives added impetus for the 
continuation of these searches to longer periods. 
We now must endeavour to continue to improve the 
precision and stability of the AAPS to be sensitive
to the 10+ year periods where analogues of the gas giants in 
our own Solar System may become detectable around other stars.
 
\section*{Acknowledgments}
The Anglo-Australian Planet Search team would like to thank 
the support of the Director of the AAO, Dr Brian Boyle, 
and the superb technical support which has been received throughout 
the programme from AAT staff -- in particular E.Penny, R.Patterson, 
D.Stafford, F.Freeman, S.Lee, J.Pogson and G.Schafer. We gratefully 
acknowledge the UK and Australian government support of the 
Anglo-Australian Telescope through their PPARC and DETYA funding
(HRAJ, CGT, AJP); NASA grant NAG5-8299 \& NSF grant 
AST95-20443 (GWM); NSF grant AST-9988087 (RPB); and
Sun Microsystems. DP is a post-doctoral researcher
at the National Fund for Scientific Research, Belgium.
This research made use of the Simbad database,
operated at CDS, Strasbourg, France.

\begin{figure}
\vspace{23cm}
\includegraphics{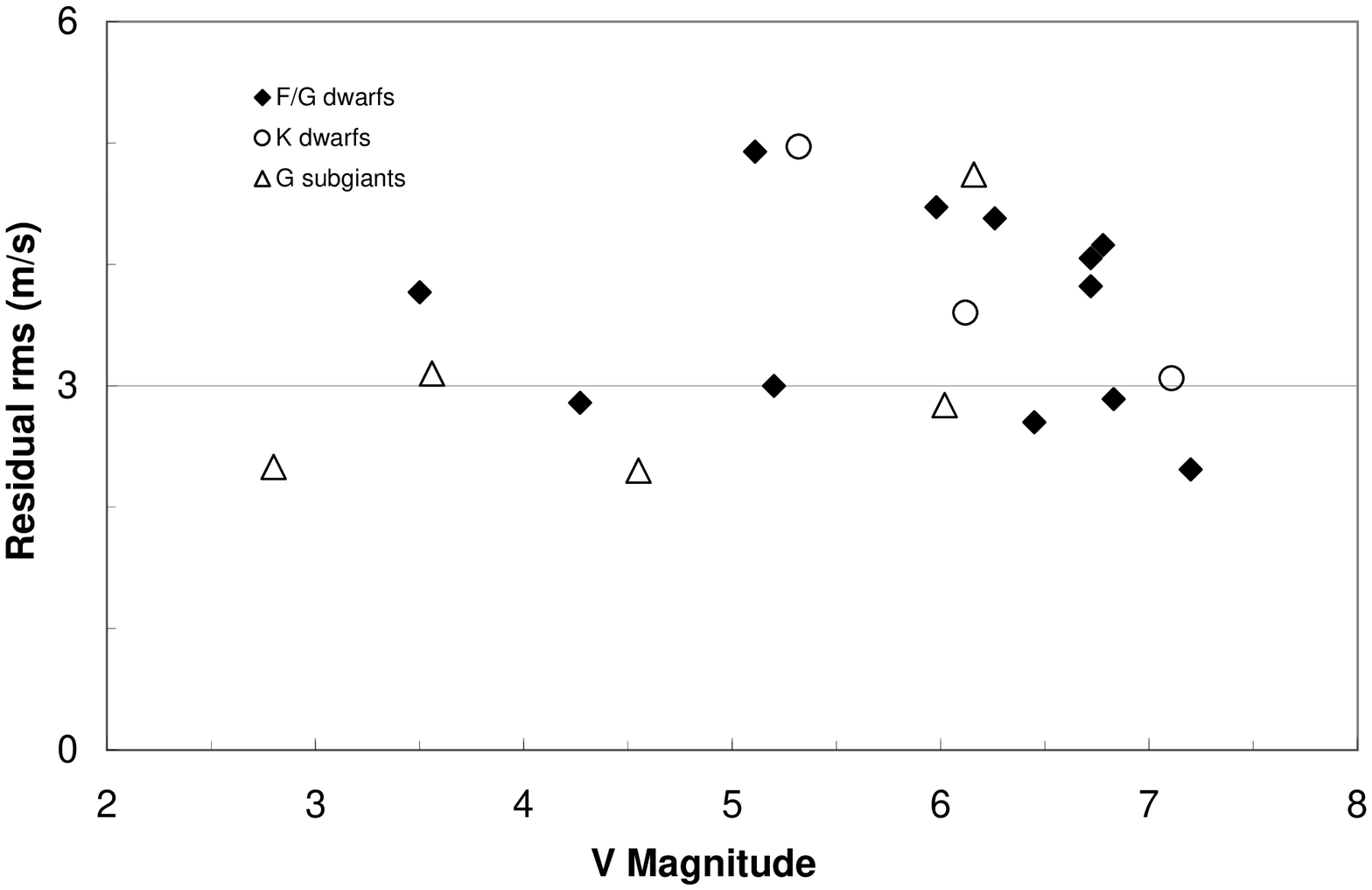}
\caption{Velocity precision as a function of V magnitude for targets reported
in Butler et al. (2001) and Tinney et al. (2001). The solid line at 
3 m~s$^{\rm -1}$ represents our target precision. Only the surveys
at Lick and Keck have demonstrated comparable precision.}
\end{figure}

\begin{figure}
\psfig{file=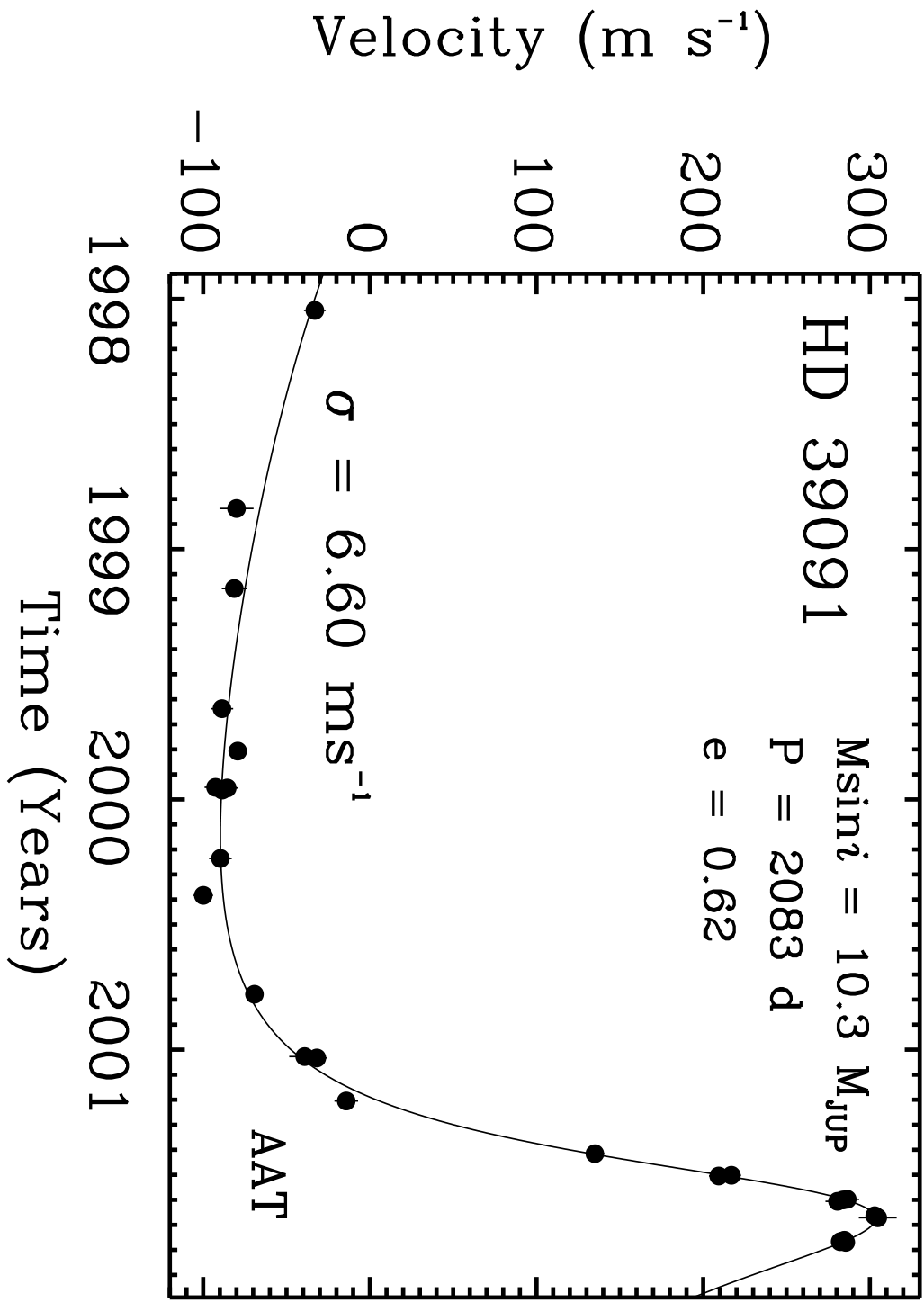,height=8.5in}
\caption{AAT Doppler velocities for HD~39091 from January 1998 to
October 2001. The solid line is a best fit Keplerian with the
parameters shown in Table 2.
The RMS of the velocities about the fit is 6.7 m~s$^{\rm -1}$ consistent
with our errors.
Assuming 1.1$\pm$0.02 M$_\odot$ (Santos et al. 2002) for the
primary,
the minimum (M~sin~$i$) mass of the companion is 10.3 M$_{\rm JUP}$ and
the semi-major axis is 3.34 au.}
\label{hd39091}
\end{figure}

\begin{figure}
\psfig{file=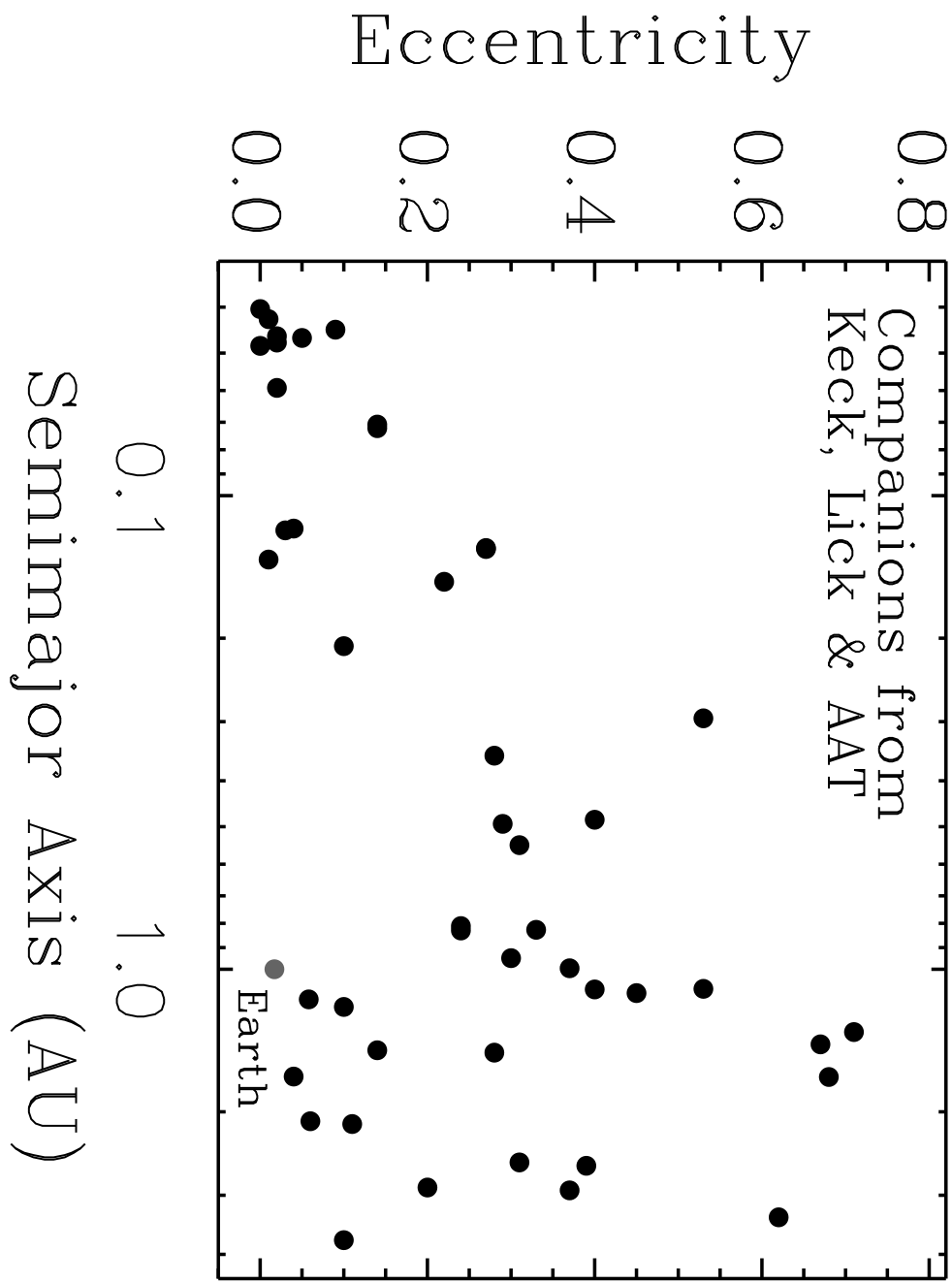,height=8.5in}
\caption{Eccentricity vs Semi-major axis with the Earth shown for
comparison. The circularity of the smallest planetary orbits
($\approx$~0.05~au) is expected due to the strong tidal
effects of the host stars. For planets orbiting between
0.15 and 3 au, eccentric orbits are the rule, not the
exception. This plot is an update of that shown in 
Butler et al. (2000) and complements the plot in fig. 13 of 
Vogt et al. (2002).
}
\label{mvse}
\end{figure}

\newpage
\begin{table}
 \centering
 \begin{minipage}{140mm}
 \caption{Velocities for HD~39091. 
Radial Velocities (RV) are referenced to the solar system barycenter 
but have an arbitrary 
zero-point determined by the radial velocity of the template.}
  \begin{tabular}{@{}lrr@{}}
JD&RV&Error \\
-2450000&m~s$^{-1}$&m~s$^{-1}$ \\
& & \\
   829.9930  &  -133.0  &  5.1 \\
  1236.0329  &  -191.4  &  6.0 \\
  1411.3249  &  -188.0  &  7.0 \\
  1473.2670  &  -173.1  &  5.2 \\
  1526.0804  &  -184.1  &  5.2 \\
  1527.0821  &  -175.7  &  4.7 \\
  1530.1280  &  -175.7  &  5.2 \\
  1629.9116  &  -181.0  &  6.0 \\
  1683.8422  &  -192.2  &  4.7 \\
  1828.1875  &  -167.9  &  4.5 \\
  1919.0989  &  -158.0  &  8.6 \\
  1921.1383  &  -143.0  &  5.2 \\
  1983.9191  &  -103.0  &  5.6 \\
  2060.8396  &    34.0  &  4.8 \\
  2092.3366  &   125.3  &  5.1 \\
  2093.3515  &   125.0  &  4.4 \\
  2127.3278  &   196.3  &  6.8 \\
  2128.3357  &   194.5  &  5.0 \\
  2130.3383  &   190.4  &  7.2 \\
  2151.2917  &   206.3  &  5.0 \\
  2154.3043  &   203.1  &  9.6 \\
  2187.1959  &   179.5  &  4.2 \\
  2188.2359  &   181.7  &  3.9 \\
  2189.2216  &   178.7  &  4.0 \\
  2190.1445  &   176.8  &  4.1 \\
\end{tabular}
\end{minipage}
\end{table}

\begin{table}
 \centering
 \begin{minipage}{140mm}
  \caption{Orbital parameters for the companion to HD~39091.}
  \begin{tabular}{@{}lr@{}}
Orbital Period & 2083 (637) \\
eccentricity & 0.62 (0.05) \\
$\omega$ & 330 (8) \\
Velocity amplitude K (m~s$^{-1}$) & 196 (18) \\
Periastron Time (JD)  &52122.1 (9.5) \\
M~sin~$i$ (M$_{\rm JUP}$) & 10.3 \\
a (au) & 3.34 \\
RMS to Fit & 6.60 \\
\end{tabular}
\end{minipage}
\end{table}

\label{lastpage}
\end{document}